\documentclass[preprint,12pt]{elsarticle}
\usepackage{graphicx}
\usepackage[hyphens]{url}
\usepackage[usenames]{color}
\usepackage{listings}
\usepackage[table,usenames,dvipsnames]{xcolor}
\usepackage{tikz}
\usepackage{hhline}
\usepackage{multirow}
\usepackage{cellspace}
\usepackage{array}
\usepackage{subfig}
\usepackage[T1]{fontenc}
\usepackage[table]{xcolor}
\usepackage{collcell}
\usepackage{hhline}
\usepackage{pgf}
\usepackage{float}
\floatstyle{plaintop}
\restylefloat{table}
\usepackage[font={footnotesize}]{caption}

\usepackage{wrapfig}
\usepackage[keeplastbox]{flushend}

\journal{Pervasive and Mobile Computing}
\begin{document}
\begin{frontmatter}

\title{Detecting Hidden Webcams with Delay-Tolerant Similarity of Simultaneous Observation}

\author[label1]{Kevin Wu}
\author[label2]{Brent Lagesse}

\address[label1]{kevinw9@uw.edu}
\address[label2]{lagesse@uw.edu}
\address{
					University of Washington Bothell\\
				 Box 358534\\
				 18115 Campus Way NE\\
				 Bothell, WA 98011-8246\\}

\begin{abstract}
	Small, low-cost, wireless cameras are becoming increasingly commonplace making surreptitious observation of people more difficult to detect.
	Previous work in detecting hidden cameras has only addressed limited environments in small spaces where the user has significant control of the environment.   
	To address this problem in a less constrained scope of environments, we introduce the concept of similarity of simultaneous observation where the user utilizes a camera (Wi-Fi camera, camera on a mobile phone or laptop) to compare timing patterns of data transmitted by potentially hidden cameras and the timing patterns that are expected from the scene that the known camera is recording.
	To analyze the patterns, we applied several similarity measures and demonstrated an accuracy of over 87\% and and F1 score of 0.88 using an efficient threshold-based classification. We used our data set to train a neural network and saw improved results with accuracy as high as 97\% and an F1 score over 0.95 for both indoors and outdoors settings.  We further extend this work against an attacker who is capable of delaying when the video is sent.  With the new approach, we see increased F1 scores above .98 for the original data and delayed data. 
	From these results, we conclude that similarity of simultaneous observation is a feasible method for detecting hidden wireless cameras that are streaming video of a user.  Our work removes significant limitations that have been put on previous detection methods.
\end{abstract}

\begin{highlights}
\item Description of the streaming video attacker model with delays.
\item A novel methodology that is able to detect hidden Wi-Fi cameras with a mobile phone.
\item The ability to defend against a delayed-transmission attacker model.
\item Evaluation of the methodology in a variety of environments and conditions.
\end{highlights}


\end{frontmatter}

\section{Introduction}

Internet connected cameras have become a pervasive feature in the world.  Most modern mobile phones contain at least one camera as do many laptops.  Additionally cheap Wi-Fi connected cameras are easy to obtain and deploy.  In addition to these devices, there are a variety of hidden cameras that are designed to evade visual detection.
The cost of obtaining and deploying such devices continues to drop as retailers such as Amazon include Surveillance Camera and Hidden Camera shopping categories that include thousands of results.  While Internet-connected cameras bring convenience to the owners, they also create security risks. Weak security mechanisms allow adversaries to exploit those IoT devices and have total control over such devices. In 2016, Mirai malware took advantage of the weak password settings of IoT devices and compromised 3.5 million devices, many of which were Wi-Fi cameras \cite{herzberg_breaking_2016}. The infected devices were located globally, including most of the countries in Europe, Asia, and North and South America \cite{_f5_2018}.  While one of the most widespread, the Mirai botnet is just one of many examples of cameras being compromised \cite{pa_pa_iotpot:_2015,krebs_hacked_nodate,fogie_abusing_2007}.  Furthermore, Wi-Fi cameras have been installed to spy on people in environments such as hotel rooms and AirBnB rentals \cite{how_2017,noauthor_yvonne_nodate,steinberg_these_nodate,polstra_am_2014}.  

Given the ease of which cameras can collect information on people without them knowing it, there is very little that has been done to detect cameras that are spying on people.  Previous work in detecting hidden cameras has generally relied on being indoors, having significant control of the environment, or performing significant manual inspection with custom hardware \cite{polstra_am_2014, lagesse_detecting_2018, roessler_hidden}.  In this paper, we extend our work \cite{wu_you_2019} in automatically detecting Wi-Fi cameras to mitigate the delayed-transmission attacker model.  The approach works both indoors and outdoors in large or small areas and can be accomplished with common computing equipment such as a mobile phone or laptop.

To address this problem, we introduce \textit{Similarity of Simultaneous Observation} to identify cameras that are streaming video of a user.  This is accomplished by utilizing a known camera in the environment such as the camera on a mobile phone and recording the environment.  Simultaneously, a networking interface enters into monitor mode and records nearby data transmissions and logs the number of bytes transmitted in each time step by each wireless device.  Next, we apply similarity measures between the data timing of the known recording and each network device.  Note that due to similarities in the size of plaintext and its resulting ciphertext when encrypted, this approach works regardless of if the camera is using encryption or is on another wireless network that we do not have credentials to join.  If the two transmissions are deemed similar enough, then we flag that device as potential webcam.  

We have evaluated our approach using over 21 hours of recordings taken from indoors and outdoors environments with varying levels of motion, resolution, and relative angles of the cameras along with a variety of traffic sources that are not observing the user in order to demonstrate the robustness of this approach.  Our experimental results show that we can achieve 100\% recall and F1 scores of 0.965 with a simple neural network and F1 scores over 0.98 with an LSTM against a more advance attack model than our original work in \cite{wu_you_2019}.

\textbf{Contributions}.   The major contributions of our work can be summarized as the following four items:
\begin{enumerate}
\item Description of a problem that has not previously been addressed in research literature in section \ref{sec:problem}.
\item A novel methodology that is able to detect hidden Wi-Fi cameras with a mobile phone in section \ref{sec:methodology}.
\item The ability to defend against a delayed-transmission attacker model in section \ref{sec:delayedattacker}.
\item Evaluation of the methodology in a variety of environments and conditions in section \ref{sec:results}.
\end{enumerate}

While the focus of our work was on streaming Wi-Fi cameras, the techniques would apply to any streaming camera as long as the system could acquire the per-time step byte counts of the device transmitting the data (for example, at a router).  
\section{Background}

Our preliminary work \cite{lagesse_detecting_2018} was the first known research to demonstrate that it is feasible to detect hidden cameras that are streaming video of a user by causing a change in the physical environment and comparing the bandwidth usage of the devices that could potentially be recording the user.  In this work, the flash on a mobile phone was used to illuminate the room, thus changing the pixels recorded by a hidden camera.  This would cause a spike in bandwidth usage.  The mobile phone uses a network card set to promiscuous mode to capture the traffic and then calculates the correlation coefficient between a vector of byte counts per time step and a vector of when the phone was flashing or not. Liu et al. \cite{liu_detecting_2018} and Cheng et al. \cite{cheng_dewicam:_2018} published similar research shortly after that also used probes to detect hidden Wi-Fi cameras.  Unfortunately, the techniques described in this work require a disturbance in the environment to operate such as rapidly flashing the flash LED on the mobile phone.  This is generally not an activity that a user would want to perform during a meeting.  Furthermore, the techniques described in these papers became increasingly ineffective in larger spaces, so it is not suitable for detecting cameras in outdoor areas or large open spaces such as shopping malls.

The reason these techniques work is due to the inter-frame video compression algorithms commonly used by Wi-Fi cameras, mobile phones, and video streaming applications. The most common modern compression algorithm used by Wi-Fi cameras, H.264, was first introduced in \cite{h.264}. One of the improvements of the H.264(MPEG-4 Part 10) is the ability to reduce the size of a video file, which requires less network bandwidth and storage space. The H.264 achieved this by removing unnecessary information, specifically, the unchanged pixels between frames. Instead, the algorithm only encodes the changing pixels with respect to reference frames. Thus, more movements occurring in the environment forced the Wi-Fi camera and the mobile phone to generate more data in network traffic and video frames.   Our system is not exclusive to H.264 and should work with any compression technique where the size of encoding at a given time is a function of the scene it is observing.

\section{Problem Statement}
\label{sec:problem}

In this section, we introduce the problem that we address in our research.  To the best of our knowledge, no previous research has directly addressed this problem.  \textbf{Given an arbitrary space, is it feasible to detect whether or not somebody is streaming video of that space.}

\subsection{System Model and Assumptions}

We assume that the user is interested in detecting a camera that is streaming video of them in an environment with a significant number of wireless networks and potentially wireless cameras.  In this paper, we refer to a scene as the area of observation recorded by a given camera.  It is not enough just to detect that a device on the network might be a camera, but also that the device is recording the scene in question.  As a result, there may be dozens of networks, dozens of streaming devices, and hundreds or thousands of total devices within range of the user.

We assume that the user has typical computing equipment available to them.  For example, they possess a computer or a mobile phone and a network card that is capable of entering into monitor mode.  We do not make explicit assumptions about whether the user is indoors or outdoors.  We do not assume knowledge of the location of the Wi-Fi camera other than that it is within range of the wireless device that is in monitor mode.  We do not assume that the user has credentials to join the network that the Wi-Fi camera is transmitting on.

\subsection{Attacker Model and Assumptions}

We make the following assumptions in this paper.  This work focuses on currently publicized attacks such as those in hotels and off-the-shelf spy cameras.  As a result, we assume the attacker lacks the motivation or technical skills to drastically reconfigure the camera.  For example, the attacker may be an AirBnB owner or even somebody who has compromised a remote webcam by guessing the password.  In this paper, we expand the attacker model beyond our previous work to include an attacker that has the ability to introduce delays in the video streaming as that caused misclassification in our previous work.

The work in this paper is designed to address 3 attacker models.  

\begin{enumerate}
	\itemsep0em 
	\item The attacker has placed a hidden camera.
	\item The attacker has compromised a device with camera capabilities.
	\item The user has deployed a device that is streaming video, but does not realize it.
\end{enumerate}

\subsection{Design Requirements}

The purpose of our work is to help users detect that a device is streaming video of them.  To this end, our work was approached with the following requirements:

\begin{itemize}
	\itemsep0em
	\item The system must work with common computing equipment that people tend to have with them most of the time.
	\item The system must work indoors or outdoors.
	\item The system must not require manipulation of the environment.
	\item The system must work even if the video is encrypted.
\end{itemize}

To the best of our knowledge, no known system or technique meets all of these requirements which has limited the effectiveness of camera detection techniques.   
\section{Methodology}
\label{sec:methodology}

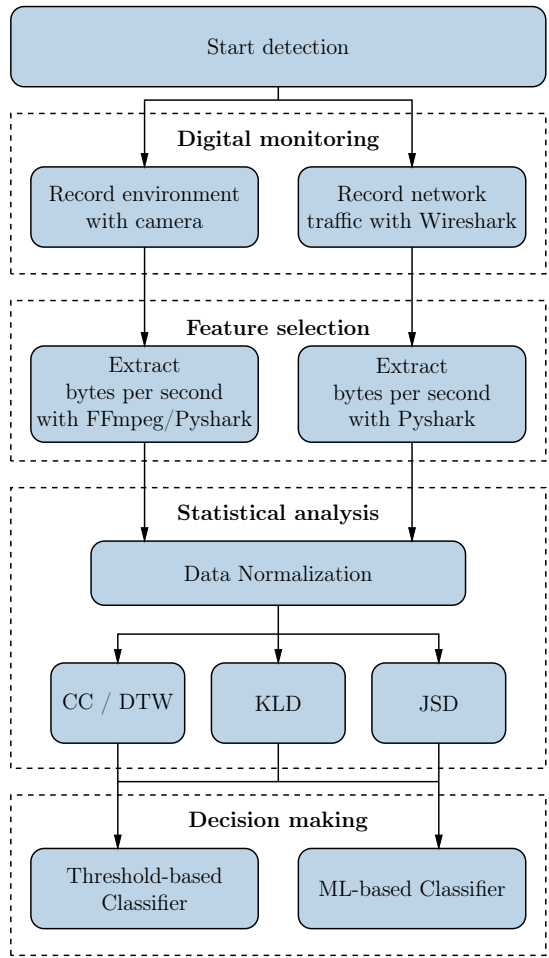
\begin{figure}
	\centering
	\renewcommand{\baselinestretch}{1}
	\scalebox{0.7}{\tikzset{every node/.style={align=center}}
\definecolor{boxshade}{rgb}{0.74, 0.83, 0.9}
\begin{tikzpicture}[scale=2.54]
\ifx\dpiclw\undefined\newdimen\dpiclw\fi
\global\def\dpicdraw{\draw[line width=\dpiclw]}
\global\def\dpicstop{;}
\dpiclw=0.8bp
\dpicdraw[fill=boxshade](4,0.2)
 ..controls (4,0.255178) and (3.955178,0.3)
 ..(3.9,0.3)
 --(0.1,0.3)
 ..controls (0.044822,0.3) and (0,0.255178)
 ..(0,0.2)
 --(0,-0.2)
 ..controls (0,-0.255178) and (0.044822,-0.3)
 ..(0.1,-0.3)
 --(3.9,-0.3)
 ..controls (3.955178,-0.3) and (4,-0.255178)
 ..(4,-0.2)
 --cycle\dpicstop
\draw (2,0) node{Start detection};
\dpicdraw[dashed](0,-1.7) rectangle (4,-0.5)\dpicstop
\draw (2,-0.7) node(C1){\bfseries Digital monitoring};
\dpicdraw[fill=boxshade](1.85,-1)
 ..controls (1.85,-0.944822) and (1.805178,-0.9)
 ..(1.75,-0.9)
 --(0.25,-0.9)
 ..controls (0.194822,-0.9) and (0.15,-0.944822)
 ..(0.15,-1)
 --(0.15,-1.4)
 ..controls (0.15,-1.455178) and (0.194822,-1.5)
 ..(0.25,-1.5)
 --(1.75,-1.5)
 ..controls (1.805178,-1.5) and (1.85,-1.455178)
 ..(1.85,-1.4)
 --cycle\dpicstop
\draw (1,-1.2) node{Record environment\\with camera};
\dpicdraw[fill=boxshade](3.85,-1)
 ..controls (3.85,-0.944822) and (3.805178,-0.9)
 ..(3.75,-0.9)
 --(2.25,-0.9)
 ..controls (2.194822,-0.9) and (2.15,-0.944822)
 ..(2.15,-1)
 --(2.15,-1.4)
 ..controls (2.15,-1.455178) and (2.194822,-1.5)
 ..(2.25,-1.5)
 --(3.75,-1.5)
 ..controls (3.805178,-1.5) and (3.85,-1.455178)
 ..(3.85,-1.4)
 --cycle\dpicstop
\draw (3,-1.2) node{Record network\\traffic with Wireshark};
\dpicdraw (2,-0.3)
 --(2,-0.4)\dpicstop
\filldraw[line width=0bp](0.975,-0.8)
 --(1,-0.9)
 --(1.025,-0.8) --cycle\dpicstop
\dpicdraw (2,-0.4)
 --(1,-0.4)
 --(1,-0.877094)\dpicstop
\filldraw[line width=0bp](2.975,-0.8)
 --(3,-0.9)
 --(3.025,-0.8) --cycle\dpicstop
\dpicdraw (2,-0.4)
 --(3,-0.4)
 --(3,-0.877094)\dpicstop
\dpicdraw[dashed](0,-3.1) rectangle (4,-1.9)\dpicstop
\draw (2,-2.1) node(C2){\bfseries Feature selection};
\dpicdraw[fill=boxshade](1.85,-2.34)
 ..controls (1.85,-2.284822) and (1.805178,-2.24)
 ..(1.75,-2.24)
 --(0.25,-2.24)
 ..controls (0.194822,-2.24) and (0.15,-2.284822)
 ..(0.15,-2.34)
 --(0.15,-2.86)
 ..controls (0.15,-2.915178) and (0.194822,-2.96)
 ..(0.25,-2.96)
 --(1.75,-2.96)
 ..controls (1.805178,-2.96) and (1.85,-2.915178)
 ..(1.85,-2.86)
 --cycle\dpicstop
\draw (1,-2.6) node{Extract\\bytes per second\\with FFmpeg/Pyshark};
\dpicdraw[fill=boxshade](3.85,-2.34)
 ..controls (3.85,-2.284822) and (3.805178,-2.24)
 ..(3.75,-2.24)
 --(2.25,-2.24)
 ..controls (2.194822,-2.24) and (2.15,-2.284822)
 ..(2.15,-2.34)
 --(2.15,-2.86)
 ..controls (2.15,-2.915178) and (2.194822,-2.96)
 ..(2.25,-2.96)
 --(3.75,-2.96)
 ..controls (3.805178,-2.96) and (3.85,-2.915178)
 ..(3.85,-2.86)
 --cycle\dpicstop
\draw (3,-2.6) node{Extract\\bytes per second\\with Pyshark};
\filldraw[line width=0bp](0.975,-2.14)
 --(1,-2.24)
 --(1.025,-2.14) --cycle\dpicstop
\dpicdraw (1,-1.5)
 --(1,-2.217094)\dpicstop
\filldraw[line width=0bp](2.975,-2.14)
 --(3,-2.24)
 --(3.025,-2.14) --cycle\dpicstop
\dpicdraw (3,-1.5)
 --(3,-2.217094)\dpicstop
\dpicdraw[dashed](0,-5.4) rectangle (4,-3.3)\dpicstop
\draw (2,-3.5) node(C3){\bfseries Statistical analysis};
\dpicdraw[fill=boxshade](3.4,-3.8)
 ..controls (3.4,-3.744822) and (3.355178,-3.7)
 ..(3.3,-3.7)
 --(0.7,-3.7)
 ..controls (0.644822,-3.7) and (0.6,-3.744822)
 ..(0.6,-3.8)
 --(0.6,-4.08)
 ..controls (0.6,-4.135178) and (0.644822,-4.18)
 ..(0.7,-4.18)
 --(3.3,-4.18)
 ..controls (3.355178,-4.18) and (3.4,-4.135178)
 ..(3.4,-4.08)
 --cycle\dpicstop
\draw (2,-3.94) node{Data Normalization};
\dpicdraw[fill=boxshade](1.3,-4.712)
 ..controls (1.3,-4.656822) and (1.255178,-4.612)
 ..(1.2,-4.612)
 --(0.4,-4.612)
 ..controls (0.344822,-4.612) and (0.3,-4.656822)
 ..(0.3,-4.712)
 --(0.3,-5.112)
 ..controls (0.3,-5.167178) and (0.344822,-5.212)
 ..(0.4,-5.212)
 --(1.2,-5.212)
 ..controls (1.255178,-5.212) and (1.3,-5.167178)
 ..(1.3,-5.112)
 --cycle\dpicstop
\draw (0.8,-4.912) node{CC / DTW};
\dpicdraw[fill=boxshade](2.5,-4.712)
 ..controls (2.5,-4.656822) and (2.455178,-4.612)
 ..(2.4,-4.612)
 --(1.6,-4.612)
 ..controls (1.544822,-4.612) and (1.5,-4.656822)
 ..(1.5,-4.712)
 --(1.5,-5.112)
 ..controls (1.5,-5.167178) and (1.544822,-5.212)
 ..(1.6,-5.212)
 --(2.4,-5.212)
 ..controls (2.455178,-5.212) and (2.5,-5.167178)
 ..(2.5,-5.112)
 --cycle\dpicstop
\draw (2,-4.912) node{KLD};
\dpicdraw[fill=boxshade](3.7,-4.712)
 ..controls (3.7,-4.656822) and (3.655178,-4.612)
 ..(3.6,-4.612)
 --(2.8,-4.612)
 ..controls (2.744822,-4.612) and (2.7,-4.656822)
 ..(2.7,-4.712)
 --(2.7,-5.112)
 ..controls (2.7,-5.167178) and (2.744822,-5.212)
 ..(2.8,-5.212)
 --(3.6,-5.212)
 ..controls (3.655178,-5.212) and (3.7,-5.167178)
 ..(3.7,-5.112)
 --cycle\dpicstop
\draw (3.2,-4.912) node{JSD};
\dpicdraw (2,-4.18)
 --(2,-4.396)\dpicstop
\filldraw[line width=0bp](0.775,-4.512)
 --(0.8,-4.612)
 --(0.825,-4.512) --cycle\dpicstop
\dpicdraw (2,-4.396)
 --(0.8,-4.396)
 --(0.8,-4.589094)\dpicstop
\filldraw[line width=0bp](1.975,-4.512)
 --(2,-4.612)
 --(2.025,-4.512) --cycle\dpicstop
\dpicdraw (2,-4.396)
 --(2,-4.396)
 --(2,-4.589094)\dpicstop
\filldraw[line width=0bp](3.175,-4.512)
 --(3.2,-4.612)
 --(3.225,-4.512) --cycle\dpicstop
\dpicdraw (2,-4.396)
 --(3.2,-4.396)
 --(3.2,-4.589094)\dpicstop
\filldraw[line width=0bp](0.975,-3.6)
 --(1,-3.7)
 --(1.025,-3.6) --cycle\dpicstop
\dpicdraw (1,-2.96)
 --(1,-3.677094)\dpicstop
\filldraw[line width=0bp](2.975,-3.6)
 --(3,-3.7)
 --(3.025,-3.6) --cycle\dpicstop
\dpicdraw (3,-2.96)
 --(3,-3.677094)\dpicstop
\dpicdraw[dashed](0,-6.8) rectangle (4,-5.6)\dpicstop
\draw (2,-5.8) node(C4){\bfseries Decision making};
\dpicdraw[fill=boxshade](1.85,-6.1)
 ..controls (1.85,-6.044822) and (1.805178,-6)
 ..(1.75,-6)
 --(0.25,-6)
 ..controls (0.194822,-6) and (0.15,-6.044822)
 ..(0.15,-6.1)
 --(0.15,-6.5)
 ..controls (0.15,-6.555178) and (0.194822,-6.6)
 ..(0.25,-6.6)
 --(1.75,-6.6)
 ..controls (1.805178,-6.6) and (1.85,-6.555178)
 ..(1.85,-6.5)
 --cycle\dpicstop
\draw (1,-6.3) node{Threshold-based\\Classifier};
\dpicdraw[fill=boxshade](3.85,-6.1)
 ..controls (3.85,-6.044822) and (3.805178,-6)
 ..(3.75,-6)
 --(2.25,-6)
 ..controls (2.194822,-6) and (2.15,-6.044822)
 ..(2.15,-6.1)
 --(2.15,-6.5)
 ..controls (2.15,-6.555178) and (2.194822,-6.6)
 ..(2.25,-6.6)
 --(3.75,-6.6)
 ..controls (3.805178,-6.6) and (3.85,-6.555178)
 ..(3.85,-6.5)
 --cycle\dpicstop
\draw (3,-6.3) node{ML-based Classifier};
\dpicdraw (0.8,-5.212)
 --(0.8,-5.5)
 --(3.2,-5.5)\dpicstop
\dpicdraw (2,-5.212)
 --(2,-5.5)\dpicstop
\filldraw[line width=0bp](3.175,-5.9)
 --(3.2,-6)
 --(3.225,-5.9) --cycle\dpicstop
\dpicdraw (3.2,-5.212)
 --(3.2,-5.977094)\dpicstop
\filldraw[line width=0bp](0.775,-5.9)
 --(0.8,-6)
 --(0.825,-5.9) --cycle\dpicstop
\dpicdraw (0.8,-5.5)
 --(0.8,-5.977094)\dpicstop
\end{tikzpicture}}
	\caption{Flowchart of the two detectors.}
	\label{framework}
\end{figure}

We propose and evaluate the detection of Wi-Fi cameras passively by recording the environment. The detection mechanism analyzes timing characteristics that exist in the recorded video and the network traffic of the Wi-Fi camera.

The default behavior of Wi-Fi cameras is based on the video compression algorithm they use. H.264, a block-oriented, motion-compensation-based video compression standard, is utilized by many modern Wi-Fi cameras and streaming applications to transfer data efficiently.  To reduce bandwidth usage, the standard only records motions between frames, in order to reduce storing overlapping information. Thus, a large amount of movement forces the Wi-Fi camera to generate and transfer large amounts of data, which creates peaks in network traffic.

The proposed framework has four major steps. The first step is to monitor the environment digitally by recording video and network traffic simultaneously. The recorded files contained timing characteristics that are essential to identify Wi-Fi camera. The second step is to extract a feature, specifically, the number of bytes per second, from both either the video file or the recorded network traffic file. This results in a vector of unsigned integers that represents each recording.  The third step is to perform statistical analysis, calculating the Pearson correlation coefficient (CC), Dynamic Time Warping (DTW) distance, Kullback-Leibler divergence (KLD), and Jensen-Shannon divergence (JSD) on the bytes-per-time step vectors. The last step is to classify each vector as belonging to a spying camera or not. Descriptions of each steps and corresponding implementation are presented in the sections below.  Figure \ref{framework} provides a visual overview of this process. 

\begin{figure}
	\centering
	\renewcommand{\baselinestretch}{1}
	\includegraphics[width=1\textwidth]{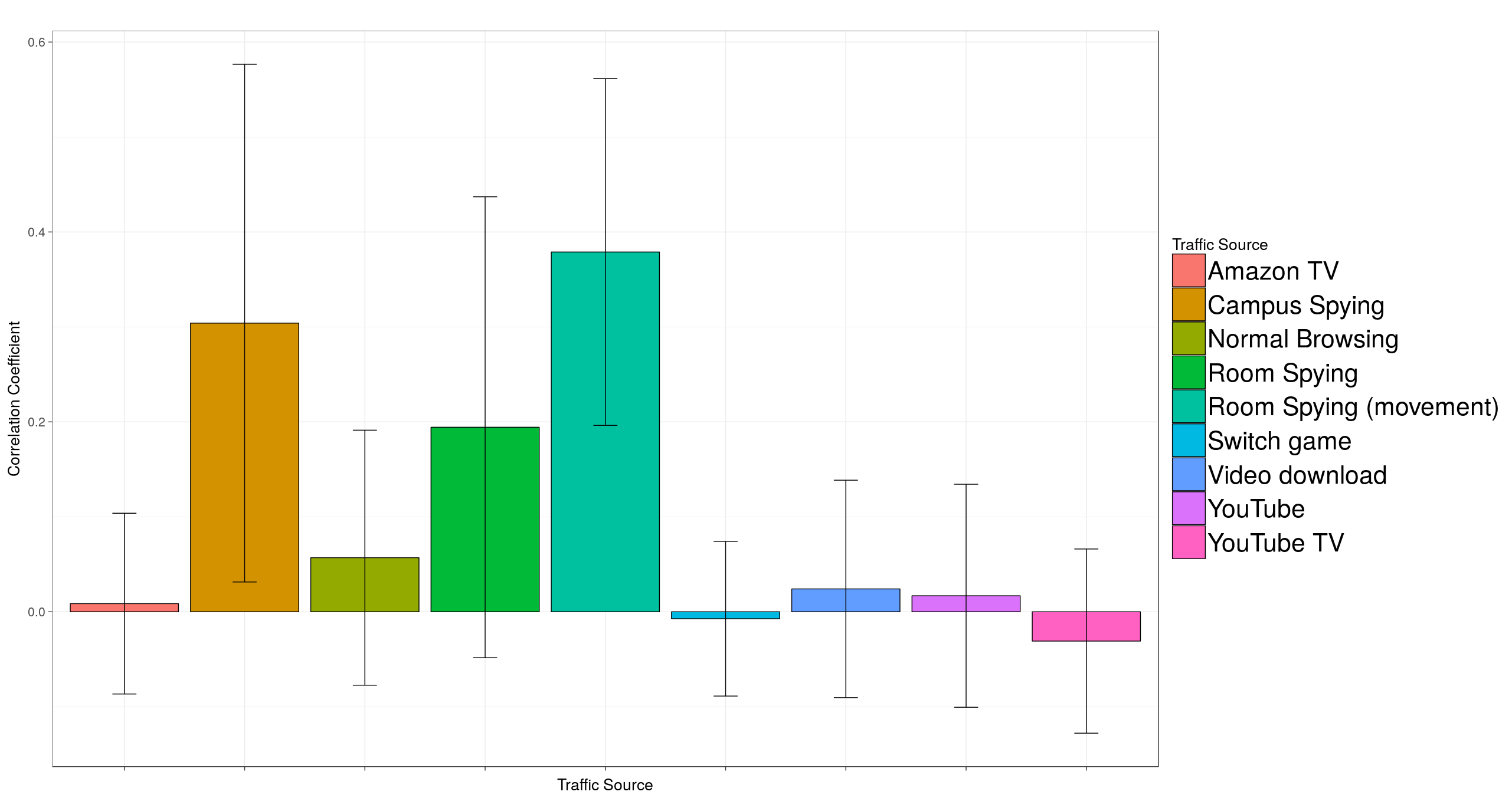}
	\caption{Correlation Coefficients for Various Traffic Sources (Error bars are one standard deviation above and below the mean)}
	\label{fig:spycamcc}
\end{figure}

\subsection{Digital Monitoring}
Digital monitoring is the first step in gathering data from the network traffic and the mobile phone. Network traffic is monitored while the mobile phone is recording the environment. In this step, the recording of the network traffic and the mobile phone are performed simultaneously.

\subsubsection{Network Monitoring}
In order to record the network traffic, a network sniffing tool is used with a network card in either promiscuous or monitor mode. Wireshark, an open source network sniffing tool supported in various platforms, is used to sniff the network traffic.  In the experiments, Wireshark is used on a Macbook Pro with macOS High Sierra 10.13.4 to perform network monitoring. The version of the Wireshark software installed on the laptop is 2.4.2 and the Network Interface Card installed on the laptop is AirPort Extreme (0x14E4, 0x170) with firmware version of Broadcom BCM43xx 1.0 (7.77.37.29.1a7).

\subsubsection{Video Recording}

To retrieve data from the environment that is monitored by the Wi-Fi camera, video recording is performed from the back camera of the mobile phone. The video recordings on the mobile phone also use a video compression algorithm to shrink the size of the video file. Mobile phones used H.264 to encode the video. This paper uses a Motorola-Z, with the OS version Android 8.0.0, to perform the experiments. The videos were recorded as either 720p or 1080p depending on the experiment, and are all in the length of one minute. The videos are encoded as MP4 files with audio support.

\subsection{Process for Features}
After the recording is completed, features are extracted from the recorded files to form data streams between IP addresses (if in promiscuous mode) or MAC addresses (if in monitor mode). Two data streams are further extracted from the recorded network traffic and the video file. While the recorded video is encoded as a MP4 file and the recorded network traffic is saved as a PCAP file, it is necessary to extract the same feature from the recorded files to perform statistical analysis. Bytes-per-time step, a shared feature in both MP4 and PCAP files, is extracted from the recordings.  Experimentally we determined that 1 second time steps provided a good trade-off between timing differences of the devices and the amount of data that the device needed to send.

\subsection{Perform Similarity Analysis}

Initially, we utilized the techniques used by \cite{lagesse_detecting_2018} to detect cyber-physical correlations; however, relying solely on Pearson's correlation coefficient resulted in an unacceptable number of false positives in some of our environments.  As shown in figure \ref{fig:spycamcc}, the correlation coefficient did result in visually different results; however, the standard deviations were so large that it was not useful as a classifier by itself.  To counter this problem, we utilized several additional distance measures.  In the case of comparing recorded videos with streaming network traffic, the correlation coefficient had so little predictive power that we did not include its results in the evaluation.

After the byte-per-second streams are extracted, we further conduct statistical analysis to calculate the relationship between the two data streams. Before performing any statistical analysis, data normalization is applied. In this project, Correlation Coefficient (CC), Dynamic Time Warping (DTW), Jensen-Shannon divergence (JSD), Kullback-Leibler divergence (KLD), Cramer distance (CD), Energy distance (ED), and Wasserstein distance (WD) are selected to measure the relationships between the two data streams.  These tests were selected because the capture a wide variety of ways that two distributions can be similar.  CC is a statistical measure to calculate the correlation between two variables and was examined due to its effectiveness in our previous work \cite{lagesse_detecting_2018}, and DTW was examined since it is used to measure similarity between two temporal sequences.  CC and DTW had the least predictive power, so we then considered other statistical measures, but we mention them here for informational purposes.  KLD calculates the differences between two normally distributed data samples and JSD measures the similarity between two probability distributions.  KLD was chosen because of the distances between the averages of spying and non-spying traffic while JSD was chosen because of the separation of the standard deviation of spying and non-spying traffic.  WD measures the underlying difference in geometries of two probability distributions.   CD is very similar to WD except that it also provides unbiased sample gradients.  ED is also similar to CD except that it is rotation invariant.  WD, CD, and ED were selected because they rely on the distance caused by the Cumulative Distribution Function which makes them more robust to minor timing mismatches caused by processing delay in the hidden camera.

\subsubsection{Data normalization}
Data normalization is performed to standardize the range of the variables in byte-per-second streams. This pre-processing step eliminates the effect of particular outliers and prevents certain objective algorithms from failing. This study utilized feature scaling to perform data normalization. Feature scaling re-scales all values in the data stream into the range between 0 and 1.

\subsection{Decision Making}
The results of the similarity analysis are used to decide whether the network stream is a Wi-Fi camera that is spying on the scene.  We examined two methods for classification.  One is a threshold-based approach where we identified values that most effectively differentiated between spying and non-spying devices.  The second is a machine learning based classifier where we trained a neural network to differentiate differentiate between spying and non-spying devices.

\subsubsection{Threshold-based approach}
The threshold selection was conducted based on the number of tests. Each collected result is further compared with the proposed threshold to determine the strength of the relationships. The threshold values are selected based on the corresponding F1 score. For each measure, we computed the F1 scores for various threshold values and selected the one with the highest F1 score.  

\subsubsection{Machine-learning-based approach}

After studying the threshold-based approach, we observed that when the system produced errors, it was usually not for all of the metrics.  Only in 24\% of our errors did we observe that all of our metrics were incorrect.   As a result, we decided to combine the metrics using supervised machine learning.   We examined a variety of machine learning algorithms and were able to achieve significantly improved results by training a neural network.

\section{Evaluation Procedure}
\label{sec:results}

In this section we evaluate the effectiveness of our approach to detecting hidden cameras in a variety of environments.  The goal of our evaluation is to understand under which circumstances the approach is effective.   We have evaluated the approach by analyzing both the network output of a Wi-Fi camera and a recording taken (but not transmitted) on a mobile phone.  We have collected data under a variety of conditions as described in table \ref{para-set} by varying the relative angle between the devices, motion in the space, resolution of the cameras, and whether the environment is indoors or outdoors.  Through these experiments we demonstrate that our work is effective in environments that prior work \cite{lagesse_detecting_2018} was not effective.

\subsection{Detectors}

We selected two likely options that a user would have to detect a streaming camera.  The first of these is to use a Wi-Fi camera and the second is to use the camera on a mobile phone or laptop.  Two Wi-Fi cameras are more likely to have stronger correlations between their network outputs due to the similarity of hardware; however, a user is more likely to carry a mobile phone than a Wi-Fi camera, so we examined both options.

\subsection{Environmental Setup}

The baseline of environment for our experiments is an 80 square meter room with lights on and with two individuals moving in space.  For reference, the results in \cite{lagesse_detecting_2018} began to significantly degrade when the device was further than 2 meters from the spying camera.  For our outdoor testing, we recorded a 250 square meter courtyard during the evening of a sunny day with one individual walking around in the space.  We also performed some experiments on a university campus with a scene that was approximately 3000 square meters (results pertaining to this environment are labeled "campus").

As seen in Table \ref{para-set}, the testing environment of the experiments is an 80 square meter room with illumination. The window size of the recordings (network traffic recording and video recording) is 60 seconds. Different angles between the hidden Wi-Fi camera and the detectors are also being considered. Testing angles included 0 degree, 90 degrees, and 180 degrees. The video compression algorithm of the Wi-Fi camera is H.264 with 720p resolution, and the video compression algorithm of the mobile phone is H.264 with both 720p and 1080p as resolutions.

\begin{table}[h]
\definecolor{header}{rgb}{0.7,0.7,0.7}
\centering
\footnotesize
 \begin{tabular}{|c|c|}\hline
 \bfseries\cellcolor{header}Parameters settings & \bfseries\cellcolor{header}Parameters Tested \\\hline\hline
 Wi-Fi camera & DCS-936L\\\hline
 Video compression & H.264/MPEG-4 \\\hline
 Mobile phone & Google Nexus 6P \\\hline
 OS platform  & Android 8.0.0 \\\hline
 Video resolution & 720p and 1080p \\\hline
 Room size  & 80 square meters \\\hline
 Courtyard size & 250 square meters \\\hline
 Illumination level of the room & Bright \\\hline
 Testing angles & $0$, $90$, and $180$ degrees \\\hline
 Window of recording & $60$ seconds \\\hline
 		\noalign{\vskip 1mm}   
 \end{tabular}
 \caption{\textsc{Parameter Settings of the Experiment.}}
 \label{para-set}
\end{table}

\subsubsection{Additional environments}

In addition to the original test data set, we have also added a new environment.  We collected data from two different cameras in a 10-square meter room with the light on. An Android-based Nexus 6P and the camera from MacBook Pro are used to perform data collection. We have collected data from different angles, including 0 degrees, 90 degrees, and 180 degrees. In total of 366 data samples have been collected. Testing Parameters of the new experiment is shown in Table \ref{new-para-set} below.

\begin{table}[h]
\definecolor{header}{rgb}{0.7,0.7,0.7}
\centering
\footnotesize
 \begin{tabular}{|c|c|}\hline
 \bfseries\cellcolor{header}Parameters settings & \bfseries\cellcolor{header}Parameters Tested \\\hline\hline
 Video compression & H.264/MPEG-4/MOV \\\hline
 Mobile phone & Google Nexus 6P \\\hline
 OS platform  & Android 8.0.0 \\\hline
 Video resolution & 720p and 1080p \\\hline
 Laptop camera & MacBook Pro \\\hline
 Video resolution & 720p \\\hline
 Room size  & 10 square meters \\\hline
 Illumination level of the room & Bright \\\hline
 Testing angles & $0$, $90$, and $180$ degrees \\\hline
 Window of recording & $60$ seconds \\\hline
 		\noalign{\vskip 1mm}   
 \end{tabular}
 \caption{\textsc{Parameter Settings from New Experiments.}}
 \label{new-para-set}
\end{table}

\subsubsection{Parameter setting} \label{param-setting}
For this research, we used an Android-based Nexus 6P and a D-Link Wi-Fi camera (DCS-936L) to perform data collection.  Network data was encrypted via WPA2. Unless otherwise noted, the parameters in Table \ref{para-set} were used for our experiments.

\subsubsection{Collected data} \label{collected-data}
In this research, we have collected in total 830 data samples from the indoors room using the Wi-Fi camera, mobile phone.  We collected 217 samples of traffic from outdoors.  We collected 260 samples of non-spying traffic.  

There is a mix of videos that capture motion and no motion.   The Wi-Fi camera recorded at 720p and observed the scene relative to the spying camera at angles of 0, 90, and 180 degrees.   The recorded video from the mobile phone included similar data except we also recorded additional data at 1080p.  

We collected videos with both the Wi-Fi camera and the mobile phone of the outdoors courtyard.   The videos were collected with and without motion.  The videos that were collected without motion were done at a time when nobody was using the courtyard.  For the motion videos, several people walked around in the courtyard while the cameras were recording.  The camera and phone were both used to record the courtyard at 0 and 90 degrees relative to the spying camera.  We also collected data from an outdoors portion of a university campus. 

For non-spying camera traffics, we collected in total 260 data samples of network traffic from Skype, YouTube, YouTube TV, Amazon TV, Switch gaming, Normal browsing, and Video downloading. Those non-spying camera traffics are used in this paper to not only produce true positives, but also avoid false positives.  We mostly focused on video-related traffic patterns, but also included non-video data for diversity.

\begin{figure}[h]
	\centering
	\renewcommand{\baselinestretch}{1}
	\subfloat[][Correlation Coefficient between Wi-Fi Camera and Spy Camera]{
		\includegraphics[width=0.48\textwidth]{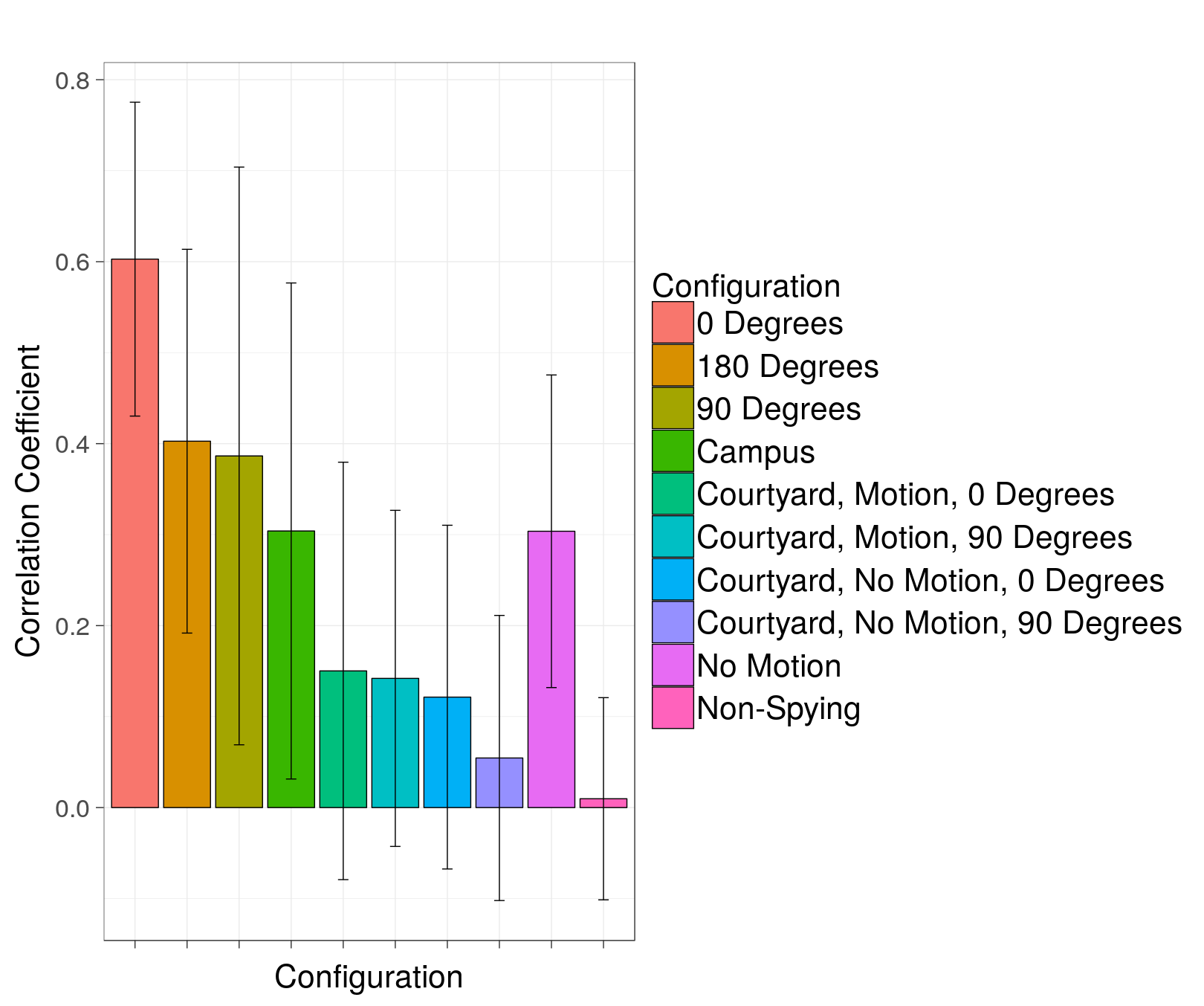}
		\label{fig:cc}

}
	~ 
	\subfloat[][JSD and KLD between Wi-Fi Camera and Spy Camera]{
		\includegraphics[width=0.48\textwidth]{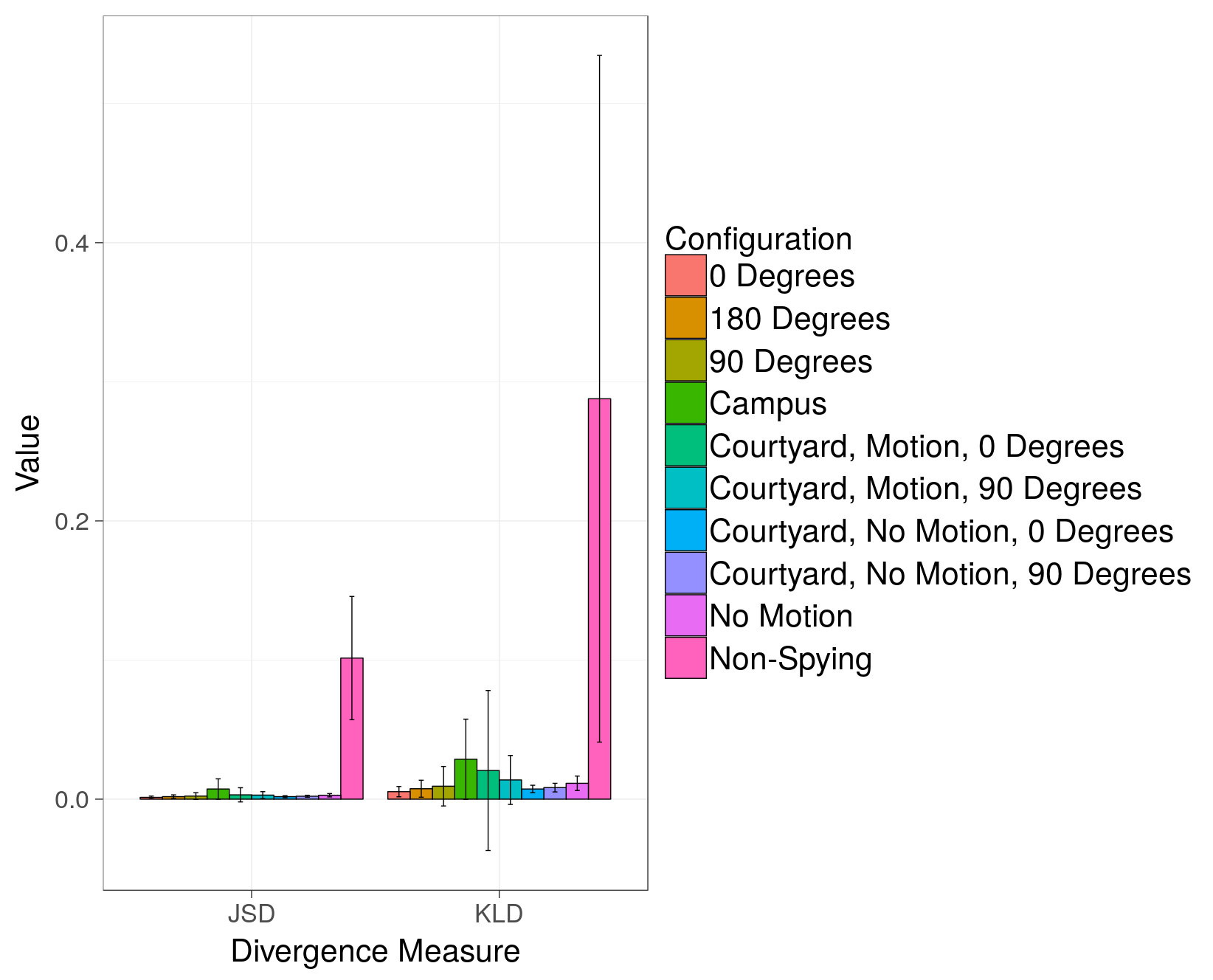}
		\label{fig:kldjsd}

}
\caption{Wi-Fi Camera Detector}
\label{fig:similarity}
\end{figure}

\begin{figure}
	\centering
	\renewcommand{\baselinestretch}{1}
\subfloat[][DTW between Mobile Phone and Spy Camera]{
		\includegraphics[width=0.48\textwidth]{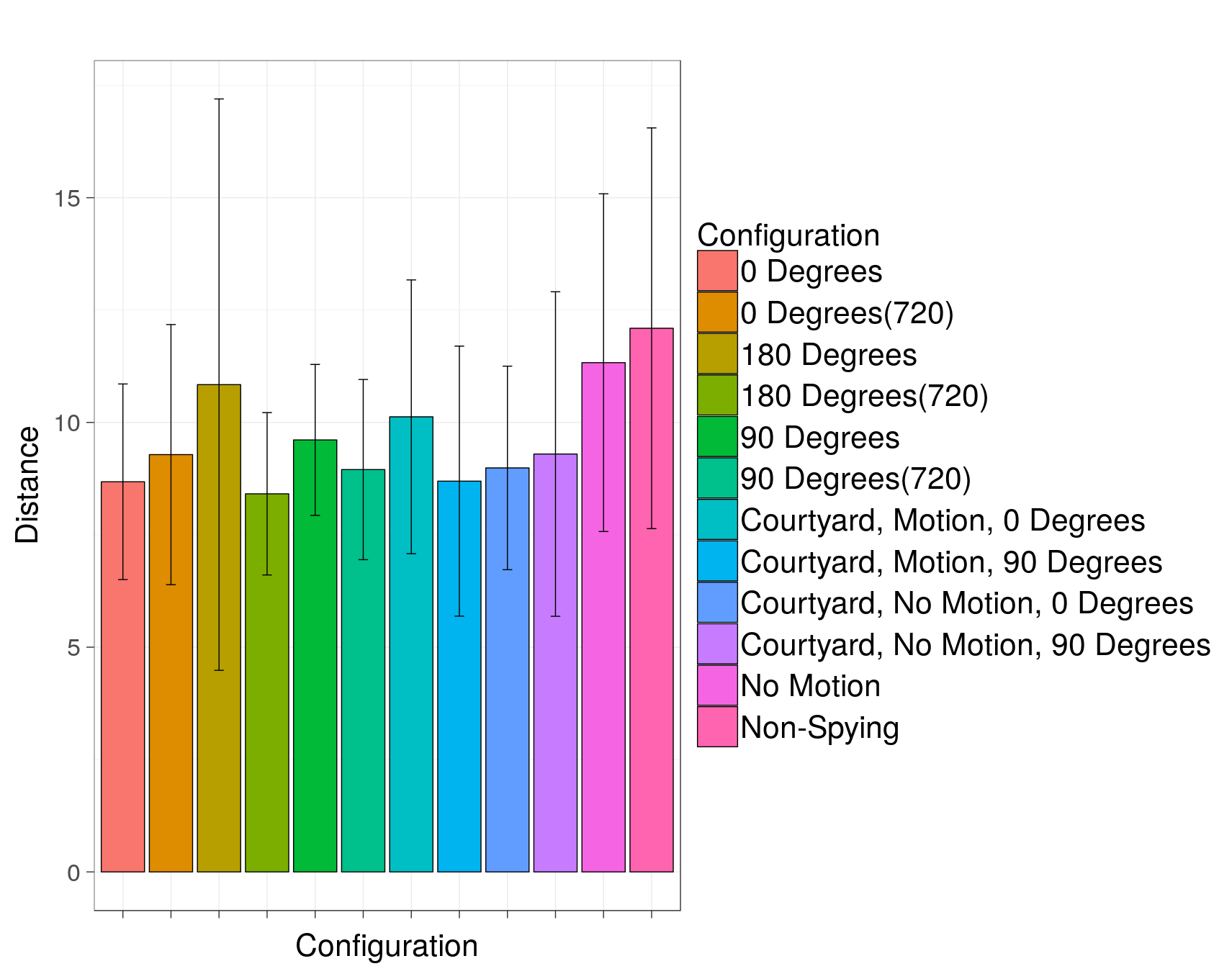}
		\label{fig:dtw}
}
	~ 
\subfloat[][JSD and KLD between Mobile Phone and Spy Camera]{
	
		\includegraphics[width=0.48\textwidth]{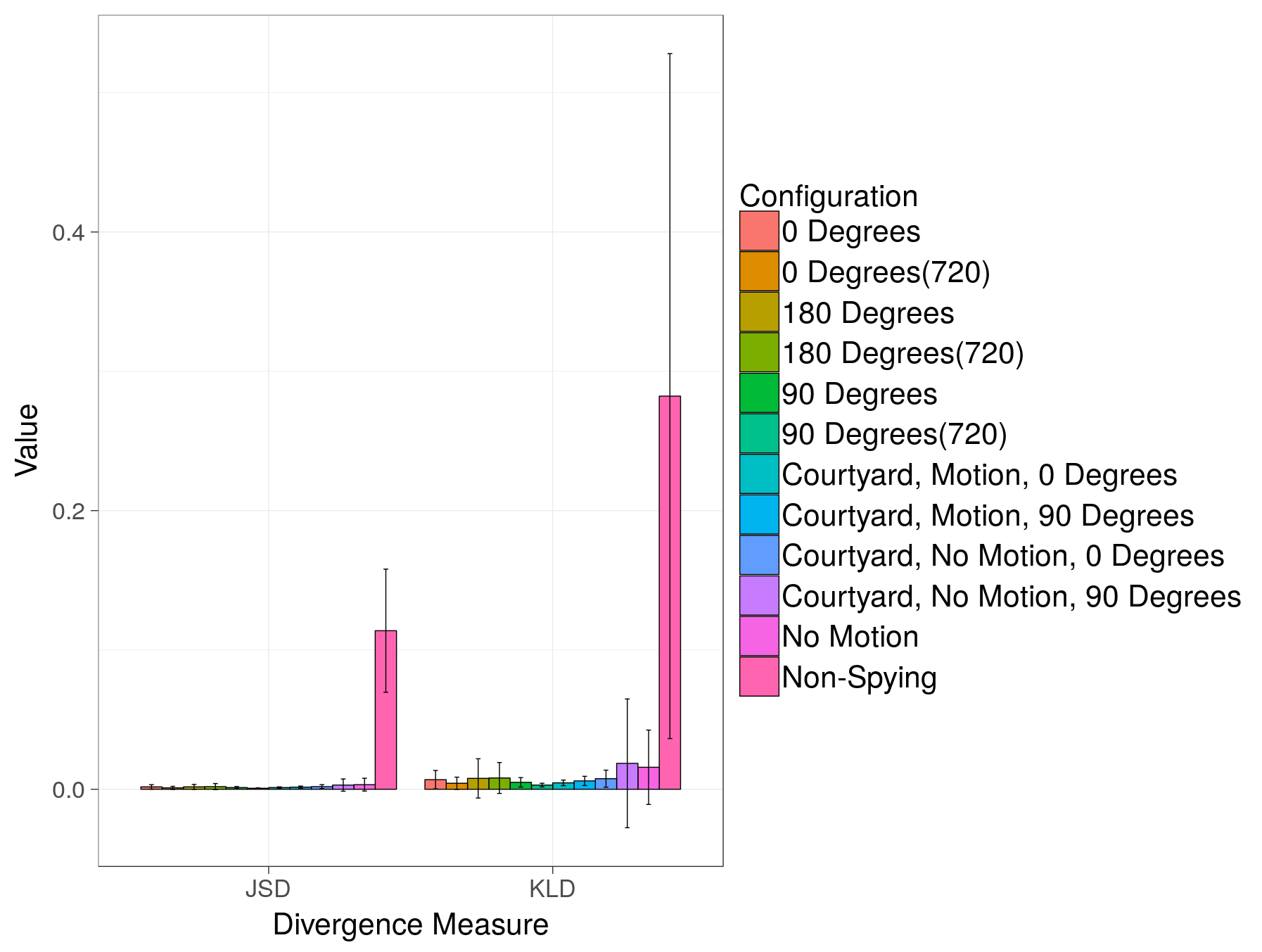}
		\label{fig:mobkldjsd}
}
\caption{Mobile Phone Detector}
	\label{fig:mobsimilarity}
\end{figure}

\subsection{Results} \label{algo-res}

In this section we present the results of the analysis of the data we collected.  These results show that the correlation coefficient measurement used in \cite{lagesse_detecting_2018} does not hold for larger outdoors spaces. They also show the added difficulty of measuring similarity between different types of devices.   From these results, utilize additional distant measures and train a neural network to assist with classification.

\subsubsection{Correlation Coefficient}

Since previous work had relied on Pearson's correlation coefficient, we first examined it as a similarity measure.   These results can be seen in figure \ref{fig:spycamcc}.  Note that while all of the situations in which there was a spying camera on average are different than the non-spying traffic, the standard deviations caused a significant overlap between spying and non-spying traffic, so we concluded that we would be unable to use only correlation coefficients for classification.   Likewise, we demonstrate in figure \ref{fig:cc} that the difference between non-spying traffic and spy cameras degrades even further when we consider results from the outdoors scenario.  

\subsubsection{Similarity Measures}

Next, we considered other measures for determining the similarity and differences between our recorded stream and the spy camera.  We examined JSD and KSD as divergence measures and showed that they provided significantly different results in spying vs non-spying traffic.   In figures \ref{fig:kldjsd} and \ref{fig:mobkldjsd} we see that for both the camera and the mobile phone, JSD has the most distance between one standard deviation above the mean for the spying video and one standard deviation below the mean for the non-spying video.  Likewise, KLD provides the largest gap between the mean of the spying video and the non-spying video.  

In our experiments between the Wi-Fi camera and the mobile phone, we noticed that there was a significant difference between the data usage of encoding on the phone and the traffic patterns of the Wi-Fi camera.  We attribute this to the low power hardware used in the Wi-Fi camera as we noticed that there were often times of significant movement where the Wi-Fi camera did not transmit any data at all and then spiked in traffic shortly after the movement.  This pattern caused the correlation coefficient to become almost useless, so we examined DTW as a distance measure.   DTW distance was only a weak predictor of whether or not a device was a spy camera as seen in figure \ref{fig:dtw}.

\begin{wraptable}{r}{0.34\textwidth}
	\begin{center}
		\begin{tabular}{ | l | c |}
			\hline
			CC & 0.21  \\ \hline
			DTW & 12.51  \\ \hline
			KLD & 0.021  \\ \hline
			JSD & 0.005  \\
			\hline
			\noalign{\vskip 1mm}   
		\end{tabular}
		\caption{\textsc{Classification Thresholds}}
		\label{tab:thresholds}
	\end{center}
\end{wraptable}

\subsubsection{Threshold-based Classifiers}\label{threshold-classifiers}

After we analyzed similarity measures as suitable for determining the distance between spying and non-spying traffic, we analyzed our results to identify optimal thresholds for classification.   The advantage of threshold classification is that it has a very low computational cost, so it has value as a classifier for low power devices.    From this analysis, we identified the best thresholds for each measure based on F1 score as shown in table \ref{tab:thresholds}.  Note that these are not necessarily always going to be the optimal threshold, but they do provide us with an understanding of an approximate starting point for a threshold-based classifier.

The results of the threshold-based classifiers can be found in table \ref{threshold}.  As expected from the analysis of distance between the means and standard deviations, KLD and JSD greatly outperformed DTW with the mobile phone detector.

\begin{table}[h]
\definecolor{header}{rgb}{0.7,0.7,0.7}
\definecolor{section1}{rgb}{0.74, 0.83, 0.9}
\definecolor{section2}{RGB}{240, 204, 176}
\definecolor{highlight}{rgb}{0.94,0.86,0.51}
\centering
\footnotesize
 \begin{tabular}{|c|c|c|c|c|c|}\hline
   \cellcolor{header}Metrics &  \cellcolor{header}F1 score & \cellcolor{header}Accuracy & \cellcolor{header}Error & \cellcolor{header}Precision & \cellcolor{header}Recall($TP$) \\\hline\hline
   \multicolumn{6}{|l|}{\cellcolor{section1}\bfseries Wi-Fi camera-based detection model} \\\hline
   CC  & $77.642$ & $77.005$ & $22.994$ & $81.159$ & $74.418$ \\\hline
   \cellcolor{highlight}KLD & $\cellcolor{highlight}88.643$ &\cellcolor{highlight} $87.165$ & \cellcolor{highlight}$12.834$ &\cellcolor{highlight} $84.384$ &\cellcolor{highlight} $93.355$ \\\hline
   JSD & $83.208$ & $84.841$ & $15.158$ & $76.497$ & $91.208$ \\\hline

   \multicolumn{6}{|l|}{\cellcolor{section2}\bfseries Mobile phone-based detection model} \\\hline
   DTW & $78.947$ & $72.173$ & $27.826$ & $67.415$ & $95.238$ \\\hline
   \cellcolor{highlight}KLD & \cellcolor{highlight}$89.185$ & \cellcolor{highlight}$87.304$ & \cellcolor{highlight}$12.695$ & \cellcolor{highlight}$83.611$ & \cellcolor{highlight}$95.555$ \\\hline
   JSD & $88.656$ & $86.782$ & $13.217$ & $83.661$ & $94.285$ \\\hline
   		\noalign{\vskip 1mm}   
 \end{tabular}
 \caption{\textsc{Threshold-based classifiers.}}
 \label{threshold}
\end{table}

\begin{table}
	\begin{center}
		\begin{tabular}{ | c | c | c |}
			\hline
			\textbf{False Positives} & \textbf{Wi-Fi Camera} & \textbf{Mobile Phone} \\ \hline
			\textbf{Total Samples}   & 61 & 135 \\ \hline
			3 & 3.28\% & 33.33\% \\ \hline
			2 & 80.33\% & 10.37\% \\ \hline
			1 & 16.39\% & 56.30\% \\
			\hline
			\noalign{\vskip 1mm}   
		\end{tabular}
		\caption{\textsc{False Positive Count}}
		\label{tab:falsepositives}
	\end{center}
\end{table}

\def\colorModel{hsb} 

\newcommand\ColCell[1]{
	\pgfmathparse{#1<50?1:0}  
	\ifnum\pgfmathresult=0\relax\bf\color{white}\fi
	\pgfmathsetmacro\compA{0}      
	\pgfmathsetmacro\compB{#1/100} 
	\pgfmathsetmacro\compC{1}      
	\edef\x{\noexpand\centering\noexpand\cellcolor[\colorModel]{\compA,\compB,\compC}}\x #1
} 
\newcolumntype{E}{>{\collectcell\ColCell}m{1cm}<{\endcollectcell}}  
\newcommand*\rot{\rotatebox{90}}

\def\colorModel{hsb} 

\newcommand\items{3}   

\begin{table}[h]
	\begin{center}
		\arrayrulecolor{white} 
		\noindent\begin{tabular}{cc*{\items}{|E}|}
			\multicolumn{1}{c}{} &\multicolumn{1}{c}{} &\multicolumn{\items}{c}{Tested} \\ \hhline{~*\items{|-}|}
			\multicolumn{1}{c}{} & 
			\multicolumn{1}{c}{} & 
			\multicolumn{1}{c}{\rot{Indoors}} & 
			\multicolumn{1}{c}{\rot{Outdoors}} & 
			\multicolumn{1}{c}{\rot{Both}} \\ \hhline{~*\items{|-}|}
			\multirow{\items}{*}{\rotatebox{90}{Trained}} 
			&Indoors  & 96.55   & 62.50  & 67.24   \\ \hhline{~*\items{|-}|}
			&Outdoors  &  81.11  & 92.31  & 83.67   \\ \hhline{~*\items{|-}|}
			&Both  &  83.02  & 84.21   & 85.71   \\ \hhline{~*\items{|-}|}
		\end{tabular}
	\end{center}
	\caption{\textsc{F1 Scores for Portability Between Indoors and Outdoors Training for Wi-Fi Camera Detector}} 
	\label{fig:CameraDetectorPortability}
\end{table}

\begin{table}[h]
	\begin{center}
		\arrayrulecolor{white} 
		\noindent\begin{tabular}{cc*{\items}{|E}|}
			\multicolumn{1}{c}{} &\multicolumn{1}{c}{} &\multicolumn{\items}{c}{Tested} \\ \hhline{~*\items{|-}|}
			\multicolumn{1}{c}{} & 
			\multicolumn{1}{c}{} & 
			\multicolumn{1}{c}{\rot{Indoors}} & 
			\multicolumn{1}{c}{\rot{Outdoors}} & 
			\multicolumn{1}{c}{\rot{Both}} \\ \hhline{~*\items{|-}|}
			\multirow{\items}{*}{\rotatebox{90}{Trained}} 
			&Indoors  & 96.55   & 73.68  & 78.79   \\ \hhline{~*\items{|-}|}
			&Outdoors  &  82.62  & 95.23  & 66.67   \\ \hhline{~*\items{|-}|}
			&Both  &  72.72  & 82.35   & 89.15   \\ \hhline{~*\items{|-}|}
		\end{tabular}
	\end{center}
	\caption{\textsc{F1 Scores for Portability between Indoors and Outdoors training for Mobile Phone Detector}} 
	\label{fig:MobileDetectorPortability}
\end{table}

\begin{figure}[h]
	\centering
	\scalebox{0.3}{\includegraphics{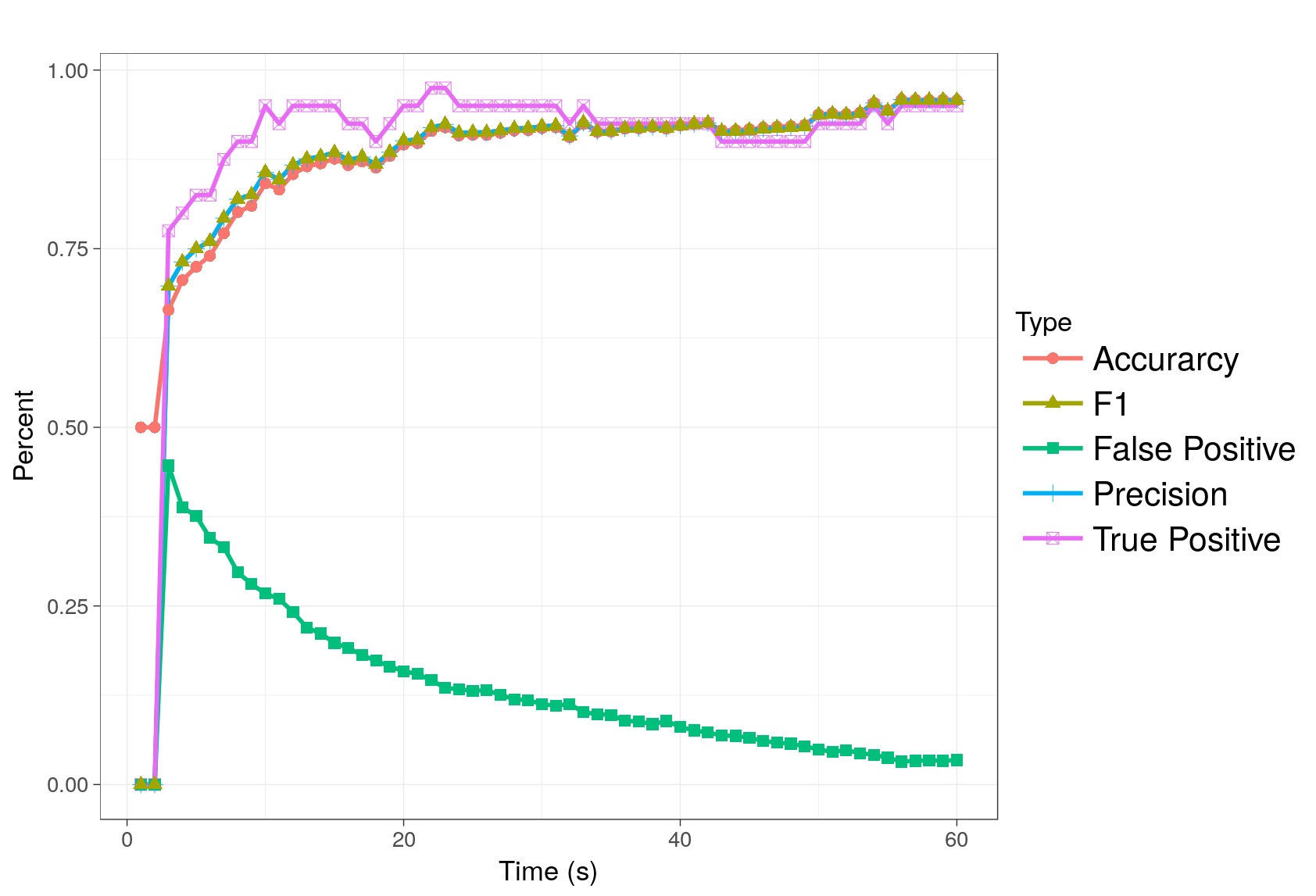}}
	\caption{Time to Convergence}
	\label{fig:convergencetime}
\end{figure}

\subsubsection{Machine Learning Classifiers}\label{ann-select}

We examined the false positives that resulted from each of the different threshold measures and noted that only 24\% of the time did all of the measures simultaneously produce a false positive.  Table \ref{tab:falsepositives} provides a breakdown of the false positives.  We hypothesized that we could utilize the lack of agreement between the similarity measures to improve our results via machine learning.  We did not examine a majority vote system because that would have only eliminated the false positives in 56\% of our samples.

\begin{table*}[h]
\definecolor{header}{rgb}{0.7,0.7,0.7}
\definecolor{section1}{rgb}{0.74, 0.83, 0.9}
\definecolor{section2}{RGB}{240, 204, 176}
\centering
\footnotesize
 \begin{tabular}{|l|r|r|r|r|r|}\hline  
   \multicolumn{1}{|c|}{\bfseries\cellcolor{header}Classifiers} & \multicolumn{1}{c|}{\bfseries\cellcolor{header}F1 score} & \multicolumn{1}{c|}{\bfseries\cellcolor{header}Accuracy} & \multicolumn{1}{c|}{\bfseries\cellcolor{header}Error} & \multicolumn{1}{c|}{\bfseries\cellcolor{header}Precision} & \multicolumn{1}{c|}{\bfseries\cellcolor{header}Recall} \\\hline\hline
   
   \multicolumn{6}{|l|}{\bfseries\cellcolor{section1}Wi-Fi camera-based detection model}     \\\hline
   Threshold-based: KLD & $88.643$ & $87.165$ & $12.834$ & $84.384$ & $93.355$ \\\hline
   Neural Network Indoors       & $96.551$ & $97.436$ & $2.564$ & $93.333$ & $100.000$  \\\hline
   Neural Network Outdoors       & $92.307$ & $94.118$ & $5.882$ & $85.714$ & $100.000$  \\\hline

   \multicolumn{6}{|l|}{\bfseries\cellcolor{section2}Mobile phone-based detection model}   \\\hline
   Threshold-based: KLD & $88.814$ & $87.453$ & $12.546$ & $81.846$ & $ 97.080$ \\\hline
   Neural Network Indoors       & $96.550$ & $96.078$ & $ 3.922$ & $93.333$ & $100.000$ \\\hline
   Neural Network Outdoors       & $95.238$ & $96.774$ & $3.226$ & $90.909$ & $100.000$  \\\hline
		\noalign{\vskip 1mm}   
 \end{tabular}
 \caption{\textsc{The Selected Best Classifiers.} } 
 \label{best-class}
\end{table*}

We examined many standard classifiers to attempt to improve above the threshold classification method.   Of these, we achieved the best performance with a neural network.  We performed grid search with 10-fold cross validation. For this study, a in total of 768 combinations of hyper-parameters are tested.  We performed the grid search separately for both the Wi-Fi camera detector and the mobile phone detector and they both produced very similar models.  The Wi-Fi camera detector's selected model had L-BFGS as the solver and the Logistic activation function. It also had three hidden layers with 13 neurons in each of them.   The only difference with the mobile phone detector was that each layer had 14 neurons.

\subsubsection{Best Classifiers}\label{final-selection}
Based on the results from sections \ref{threshold-classifiers} and \ref{ann-select}, we selected the best threshold-based and machine-learning-based classifiers for the two detection models. The selected best classifiers are presented in Table \ref{best-class} below.

Table \ref{best-class} presents the best classifiers for the two detection models. As seen in the table, neural network models outperformed threshold-based classifiers both in terms of the F1 score and an accuracy rate achieving above 94\%. Moreover, both of the neural network models had a 100\% recall rate, so scoring measurements that focus more heavily on True Positives would result in even better scores.

\subsubsection{Convergence Time}

While all of the tests described in this paper were run on 60 seconds of observation, we also examined the convergence rate of detection.  We randomly selected 1 spying camera device and 69 non-spying camera devices then analyzed our results at each time step.   Figure \ref{fig:convergencetime} shows that our results when averaged over 40 trials.  Generally the spying camera is identified within a 10 seconds, and the rest of the time is spent weeding out the false positives.  We see that the F1 score exceeds 0.90 within 20 seconds.

\subsubsection{Model Portability}

In this portion of the evaluation we examined the portability of the models between indoors and outdoors spaces.   Figures \ref{fig:CameraDetectorPortability} and \ref{fig:MobileDetectorPortability} present a matrix summary of the results by showing the F1 scores for our models when the data is partitioned into Indoors, Outdoors, and Both and then the model is trained and tested on samples from each set.  From these results, we see that, as one would expect, the best results are achieved when the model is trained only with the class of data that it will be used to test with.  We also note that training with the outdoor data provided much better results for non-outdoors testing than occurred with indoor training data.  In general we conclude that it is best to use separate models for drastically different types of space, but even if you use a combined model, there will still be value to the results.

\subsubsection{Performance of the ANN}

We further examined the performance of the selected model in the new environment. The best ANN model was selected  and trained with the old dataset, then made predictions on the new dataset we collected. The results of the predictions on the new dataset shown in Table \ref{new-model-comparison}. As shown in the table, the performance of the model suffered as the accuracy rate drops to 81.94\%. The overall performance also drops while numbers of false positive and false negative increases. It is stated that different environment settings did affected the magnitudes of the changes pixels in the H.264 encoding thus degrades the performance of the model. 

In order to boost the performance of the model, we introduced several new features into the ANN model. Cramer distance, Energy distance, and Wasserstein distance are further implemented into the new ANN model. Performance of the new ANN model is also presented in Table \ref{new-model-comparison}. Compared with the previous model, the model with new features is able to boost the recall rate to 93.37\%. The new model trained with the new features can classify more true positives than the previous model. However, the accuracy rate and overall F1 score stays around the same ranged between 82\% to 84\%. It is also noticed that the Precision rate decreased by 4\%, which indicates the new model classified slightly more false positives. 

To prevent overfitting the model, we examined the performance of the model by reversed the training dataset with the testing dataset. By training the model with the new dataset and made predictions on the old dataset, we can observed the sustainability of the model’s performance in different environments. As shown in table \ref{new-model-comparison}, The model performs about the same as the previous model. Reversing the train-test dataset gave us slightly better results as all of the performance metrics improved slightly. The accuracy rate increased by 3\% as well as the overall F1 score. The persistence of the performance has proved that the model could serve as a uniformed detection classifier that can make predictions in other environments with promising results.

\begin{table*}[h]
\definecolor{header}{rgb}{0.7,0.7,0.7}
\definecolor{section1}{rgb}{0.74, 0.83, 0.9}
\definecolor{section2}{RGB}{240, 204, 176}
\centering
\footnotesize
 \begin{tabular}{|l|r|r|r|r|r|}\hline  
   \multicolumn{1}{|c|}{\bfseries\cellcolor{header}Classifiers} & \multicolumn{1}{c|}{\bfseries\cellcolor{header}F1 score} & \multicolumn{1}{c|}{\bfseries\cellcolor{header}Accuracy} & \multicolumn{1}{c|}{\bfseries\cellcolor{header}Error} & \multicolumn{1}{c|}{\bfseries\cellcolor{header}Precision} & \multicolumn{1}{c|}{\bfseries\cellcolor{header}Recall} \\\hline\hline

   ANN Indoors & $96.55$ & $96.08$ & $ 3.92$ & $93.33$ & $100.00$ \\\hline
   ANN: New environment & $83.30$ & $81.94$ & $ 18.06$ & $80.94$ & $85.82$ \\\hline
   ANN: New features & $84.08$ & $82.32$ & $ 17.68$ & $76.47$ & $93.37$ \\\hline
   ANN: Reversed train-test & $87.34$ & $85.47$ & $ 14.53$ & $80.54$ & $95.39$ \\\hline

		\noalign{\vskip 1mm}   
 \end{tabular}
 \caption{\textsc{Performance comparison of the new models.} } 
 \label{new-model-comparison}
\end{table*}

\subsubsection{Video Delays}
\label{sec:delayedattacker}

\begin{figure}[h]
	\centering
	\renewcommand{\baselinestretch}{1}
	\includegraphics[width=1\textwidth]{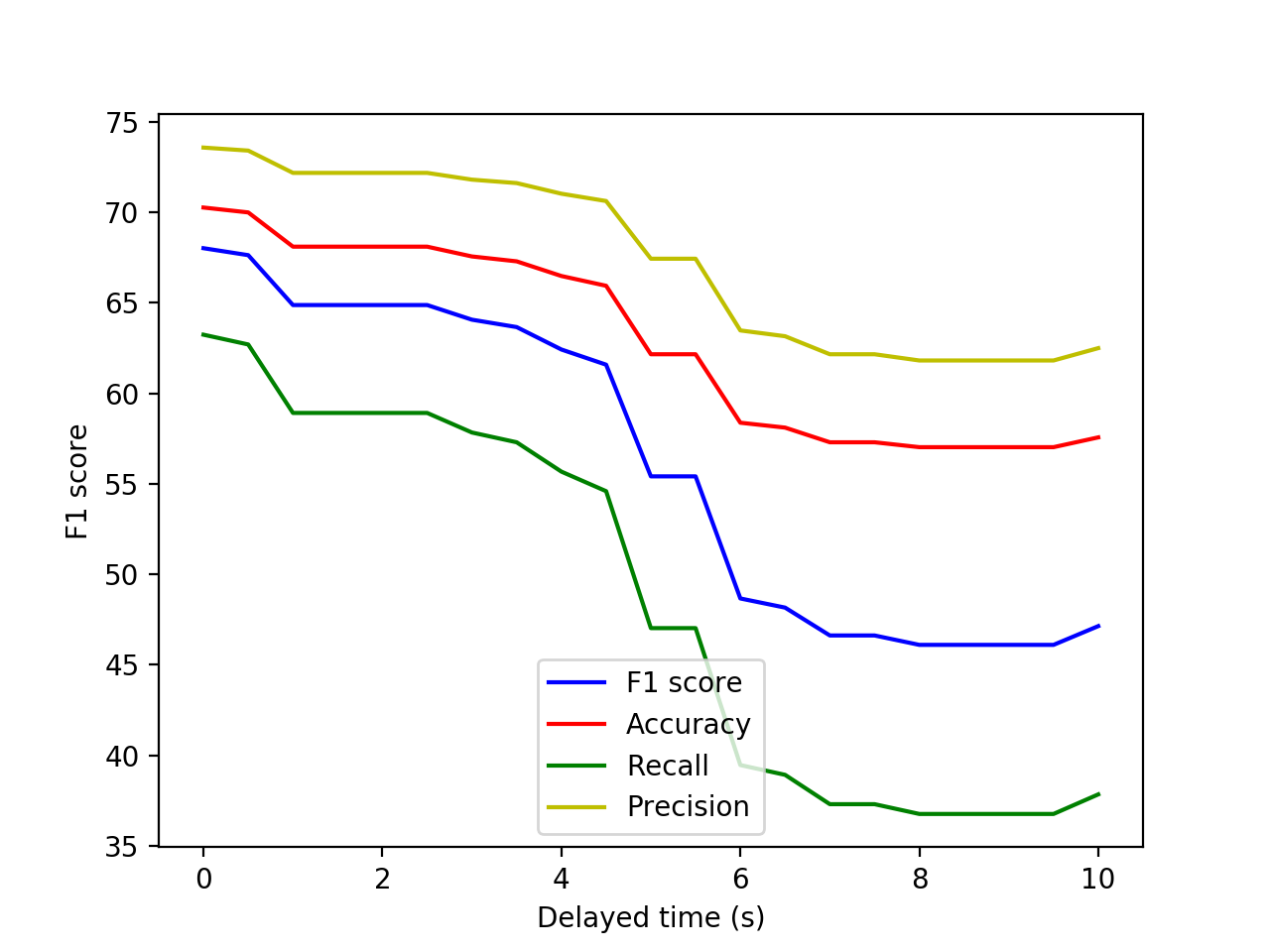}
	\caption{Effect of Delay Attack on Original Model}
	\label{fig:delayAttack}
\end{figure}

The adversaries might perform delay viewing attacks on the hidden camera internet stream to counter the detection. Delay viewing attack is carried out by the intentionally delay the internet stream for several seconds to differentiate it from normal internet stream of a Wi-Fi camera. Addressing the possibility of the delay viewing attack on the hidden Wi-Fi camera, we compare and evaluate the performance of the old ANN model and the new ANN model. The results of the performance of the old model is presented in Figure \ref{fig:delayAttack}.

As shown in Figure \ref{fig:delayAttack}, the old model did not perform well against the delay viewing attack. The accuracy rate stays around 67\% within 5 seconds of delay then drops dramatically below 60\% for longer delay. The precision rate is always higher than the recall rate indicates that the model predicts more false negatives than false positives. This has shown that the model was falsing predicting hidden camera stream as other internet streams. The performance of the model stays mediocre until the delay reached over 5 seconds. Recall rate even drops below 40\% if the delay is longer than 6 seconds. It has shown that the old model did not have the ability to detect hidden cameras under the delay viewing attack.  

\begin{figure}[h]
	\centering
	\renewcommand{\baselinestretch}{1}
	\includegraphics[width=1\textwidth]{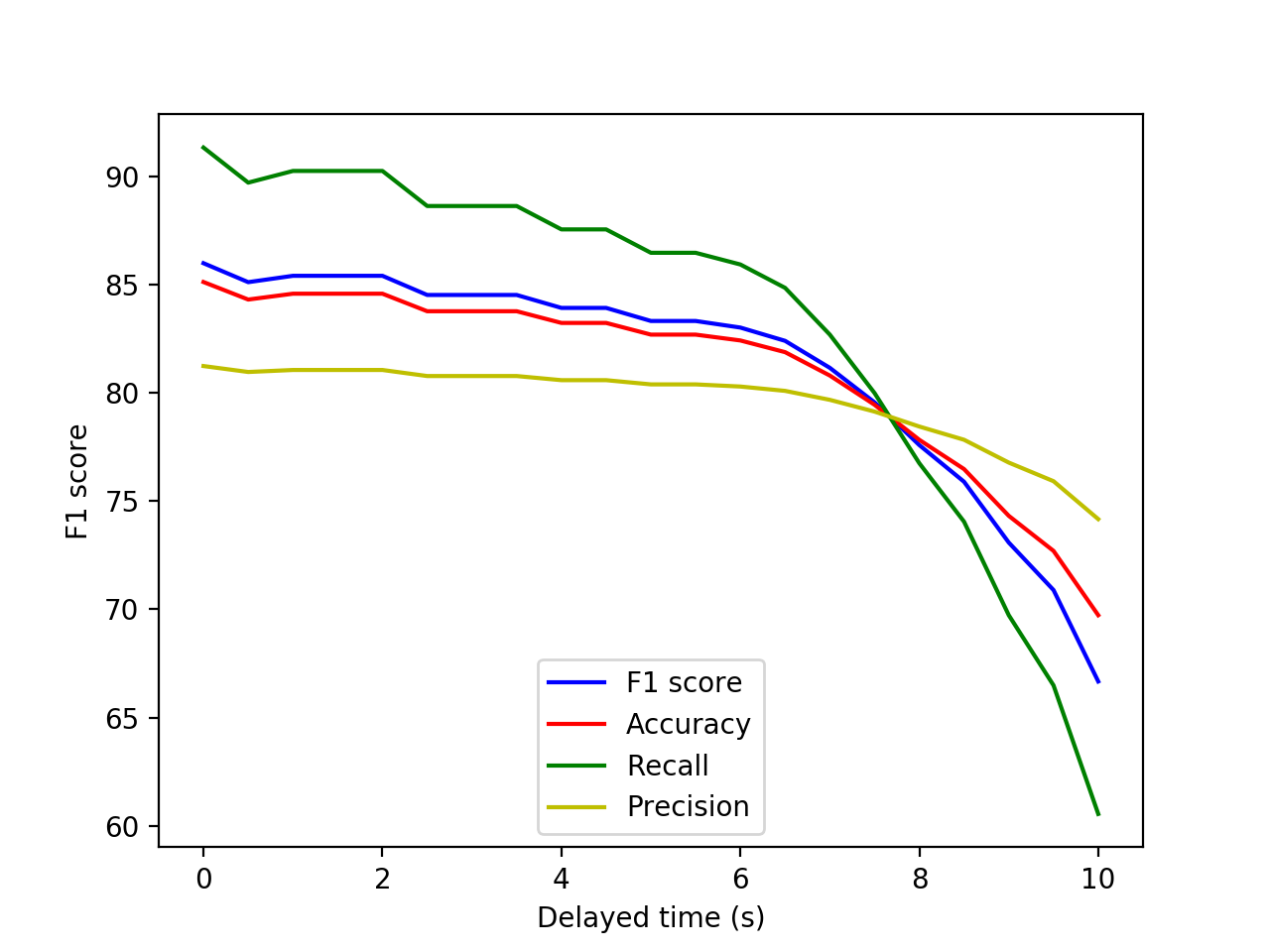}
	\caption{Effect of Delay Attack on Model with Improved Metrics}
	\label{fig:delayAttackNewMetrics}
\end{figure}

The performance of the new model under delay viewing attack is shown in Figure \ref{fig:delayAttackNewMetrics}. As shown from the figure, the new model performs better than the old model while it maintained an accuracy rate around 83\% within 5 seconds of delay then slightly decreased to 70\% for longer delay. The new model performs opposite as the older model by having a higher recall rate than the precision rate. It is showing that the model predicts more false positive than false negative. The new model is more feasible than the older model since it had less false negatives, indicates that it can predict more hidden cameras rather labeled them as other internet streams. The recall rate maintained higher than precision rate and was higher than 80\% until delay reach around 7.8 seconds. It is showing that the model had the ability to predict sufficient amount of true positives until the delay is longer than 8 seconds. The performance of the model drops dramatically over 8 seconds of delay and the accuracy rate maintained over 70\% for 10 seconds delay. Although the performance of the new model degrades, it still gave us promising results for any delay attack shorter than 8 seconds. As a result, the new model served as a better classifier than the older model to counter delay viewing attack.

\subsubsection{LSTM-based Approach}

In order to achieve better predicting results, we implemented an LSTM model. The LSTM model is a well-known artificial recurrent neural network which predicts time series data. The selected LSTM model had 50 neurons in the first hidden layer, with 1 neuron in the output layer with the sigmoid function. It also utilized an Adam optimizer and the selected batch size is set to 48. A stepped function is applied after the output layer. A threshold of 0.00001 is applied within the stepped function. The LSTM model is trained with the old dataset and make predictions based on the new dataset. The performance of the LSTM model is shown in Table \ref{LSTM}. As shown in the table, the LSTM model performs well on predicting time-series data. The accuracy rate had achieved 99.72\% and the overall F1 score reached 99.76\%. Moreover, the model is able to eliminate all the false positives while the recall rate achieved 100\%. It shows that the model is perfectly fit for detecting all hidden cameras without mispredict any hidden camera stream. 

To prevent overfitting the model, we also examined the performance of the LSTM model by reversing the training dataset with the testing dataset. As shown in table \ref{LSTM}, The model performs about the same before reversing the train-test dataset. Both the accuracy rate and the overall F1 score drops around 1\%, and the recall rate drops around 2.5\%. It indicates that reversing the train-test dataset made the model allowed more false negatives. Besides the slight downgrade of the evaluation metrics, the overall performance of the model stays excellent as the accuracy rate of 98.61\% and overall F1 score of 98.66\%. It indeed showed that the LSTM model is perfectly fit for detecting hidden Wi-Fi cameras, moreover, it is a uniformed classifier that can predict hidden cameras in different environment settings. 

To address the possibility of delay-viewing attack on the hidden Wi-Fi camera, we further evaluate the performance of the LSTM model on delayed dataset. We have evaluated the performance of the model on two different scenarios: delay viewing with 5 seconds and delay viewing with 30 seconds. As shown in table \ref{LSTM}, The LSTM model still performs well for both of the delaying scenarios. Both of the accuracy rate and the overall F1 score stayed between 97\% and 98\% while the error rate increased slightly by 2\%. It has shown that the LSTM model had the ability to counter delay-viewing attack even with 30 seconds of delay. 

\begin{table*}[h]
\definecolor{header}{rgb}{0.7,0.7,0.7}
\definecolor{section1}{rgb}{0.74, 0.83, 0.9}
\definecolor{section2}{RGB}{240, 204, 176}
\centering
\footnotesize
 \begin{tabular}{|l|r|r|r|r|r|}\hline  
   \multicolumn{1}{|c|}{\bfseries\cellcolor{header}Classifiers} & \multicolumn{1}{c|}{\bfseries\cellcolor{header}F1 score} & \multicolumn{1}{c|}{\bfseries\cellcolor{header}Accuracy} & \multicolumn{1}{c|}{\bfseries\cellcolor{header}Error} & \multicolumn{1}{c|}{\bfseries\cellcolor{header}Precision} & \multicolumn{1}{c|}{\bfseries\cellcolor{header}Recall} \\\hline\hline

   LSTM model & $99.76$ & $99.72$ & $ 0.28$ & $99.53$ & $100.00$ \\\hline
   LSTM: Reversed train-test & $98.66$ & $98.61$ & $ 1.39$ & $99.90$ & $97.44$ \\\hline

   LSTM: Delay 5 sec. & $98.91$ & $98.07$ & $ 1.93$ & $100.00$ & $97.85$ \\\hline
   LSTM: Delay 30 sec. & $98.89$ & $97.80$ & $ 2.20$ & $97.80$ & $100.00$ \\\hline

		\noalign{\vskip 1mm}   
 \end{tabular}
 \caption{\textsc{Performance of the LSTM models.} } 
 \label{LSTM}
\end{table*}

\section{Discussion}

The results we obtained in this study demonstrate that there are 4 main points of concern for determining how accurately one can detect hidden cameras using the passive approaches described in this paper.  These include the changes in the physical world that can be observed by the devices, the fidelity of the camera, the network transmission, and the background traffic from other devices.   In other words, to theoretically predict your results, you need to answer the following questions:  i) What is happening in the physical world? ii) How is it being recorded? iii) How is it being transmitted? iv) How is it different from other transmissions?

\subsection{Scene Change}
\label{sec:scenechange}

Scene change describes the scene that the cameras are recording.  To demonstrate this point, consider two cameras that are facing each other with a television in between them.   The camera facing the front of the television would record significant change whereas the one facing the back would record no change.  The primary variables that can affect detection are the relative placement of the cameras which affects the portion of overlap of the recorded scene, and the magnitude of the movement in the overlap of the recorded scene.  The placement of the cameras affect the detection since their location affects the number of pixels that are simultaneously altered due to a change in the scene between shared between two recordings.  The magnitude of the movement in the scene affects the detection since no movement or constant movement will be easy to confuse with periodic network traffic that has a similar transmission frequency to the I-Frame transmission frequency for the codec or for near constant bitrate traffic, respectively.

\subsection{Camera Fidelity}

Camera fidelity describes the quality of the recording made by the camera.  To demonstrate this point, consider an extreme case where the camera only records a single pixel that is either black or white vs a camera with 1920x1080 resolution.  The higher resolution camera would be able to pick up subtle changes whereas the 1 pixel camera would not be able to do so.  The primary variables that can affect detection are the resolution of the camera, the video codec, and the optics of the camera.  The resolution affects the number of pixels that a change in the scene affects;  normalization can mask this in some cases, but not when a particular movement fails to register a change in lower resolution cameras.  The video codec and its associated parameters can affect how many pixels are reported as changed especially depending on the compression technique.  The optics of the camera can affect how sensitive a camera is to change and whether or not minor changes are detected. 

\subsection{Network Transmission}

Network transmission describes how the data is disseminated by the camera.  To demonstrate this point, consider a camera that is streaming over TCP and a camera that is streaming over UDP.  Congestion in the network could cause the TCP camera to back off and modify its transmission speed whereas the UDP camera would transmit as fast as data was available, so the exact same scene could appear on the network with different bandwidth consumption.  The primary variables that can affect detection are transmission delays, differing protocols, and the differing parameters used even when the protocols are the same.  The delay can be due to processing delay because of low-power computing hardware, a phenomenon we experienced in our experiments, or due to customization by the attacker to try to evade detection.  As mentioned before, different protocols for transmitting data can affect the timing and quantity of data transmitted.  Furthermore, some protocols that adapt to bandwidth availability can cause issues if they adapt during the middle of bandwidth sampling since it would throw off our normalization process.  Similarly, each network transmission protocol can be configured with different parameters that could result in different timings or bandwidth usage patterns.

\subsection{Background Traffic}

Background traffic describes the network traffic that is being transmitted by devices other than the spy camera.  Since the usefulness of detecting spy cameras depends on being able to differentiate between the spy camera and other network devices, devices that have transmission patterns similar to the timing of movement in the recorded scene will result in false positives as mentioned in section \ref{sec:scenechange}.

\subsection{Limitations}

If an attacker switches from an interframe compression algorithm such as H.264 to an intraframe or constant bit rate compression algorithm then our technique will be ineffective at detecting that camera; however, this switch comes with a cost of increased bandwidth usage.   While many cameras still support MJPEG our experience has been that the cameras we have evaluated default to H.264 and some of them no longer include MJPEG support.  We could extend our approach to also record using MJPEG and look for correlations since JPEG will compress each frame differently based on the colors in the scene.

An attacker could also modify the software running on the webcam to inject additional signals into the data transmission rates that are not as expensive as a CBR codec would be.  The topic of how an attacker can optimize this injection is a topic for future work.  We would need to augment our system with approaches such as network anomaly or protocol detection to be able to detect such an attack.

Additionally, we are limited to streaming cameras with this approach.  As future work we are examining improved techniques for detecting cameras that are not streaming data.  Currently, this approach would need to be used as one technique in an anti-spying toolkit.

\section{Related Work}
 
Related research has focused on identifying services, applications, websites, and connected devices with various detecting mechanisms. Since network traffic contained critical information regarding communicating entities and ongoing communications, most of the research concentrated on detecting targets by utilizing the data embedded within network traffic. Some studies introduced in perform timing analysis is also related to our work.

\subsection{Network traffic analysis}

Geer et al. \cite{geers_core_2017} demonstrate that network traffic analysis is a powerful tool to identify targets regarding of the network traffic volume that is generated daily. Their research included several features of the network traffic, such as frequency, volume, and timing, that are favorable for the attackers to identify particular patterns. Moreover, encryption over network traffic does not prevent adversaries from studying those features. The findings allowed adversaries to identify certain behavior and services from the network traffic. Coull et al. \cite{coull_traffic_2014} researched network traffic analysis for Apple iMessage. The study looked into the volume of the encrypted network traffic that is being transferred and found that adversaries can successfully learn the victim's actions, language used, and the length of the messages with 96\% of accuracy. 

Siby et al. \cite{siby_iotscanner:_2017} focused on an IoT-rich environment and privacy concerns.  They discovered existing wireless infrastructure by analyzing the numbers of Frames, mFrames, cFrames, and dFrames; network traffic volume; and send-to-received ratio passively identify IoT devices.   Gong el at. \cite{gong_website_2011} studied the feasibility of Dynamic Time Warping (DTW) on network traffic patterns. The study showed that website fingerprinting is applicable, even with noisy network traffic, by applying DTW with traffic analysis.

\subsection{Timing analysis}

Feghhi et al. \cite{feghhi_time_2017} researched the effectiveness of timing-based attacks against encrypted network traffic and were able to infer web pages more than 87\% of the time. Other studies have demonstrated that performing timing analysis reveals victim nodes within anonymizing systems \cite{murdoch_low-cost_2005, shmatikov_timing_2006}. 

Apthorpe et al. \cite{apthorpe_smart_2017} performed experiments on IoT smart home devices. They discovered that the network traffic of those devices often revealed potential information about user interactions. Based on the sending/receiving rates of the streams, they were able to map live traffic to user behaviors. This research indicates that the network streams of IoT devices have certain attributes that are controllable by the users. We expect to adapt their findings to build a novel IoT sensor detection method based on certain movement interactions. A timing analysis on a low-latency network has also been discussed \cite{murdoch_low-cost_2005,shmatikov_timing_2006}. Both studies have pointed out that the timing characteristics of network traffic tend to be remained. We intend to extend their findings to perform statistical analysis on the timing characteristics of Wi-Fi cameras.

\section{Conclusion}

This paper has proposed and evaluated a novel method, \textit{Similarity of Simultaneous Observation}, for detecting streaming Wi-Fi cameras.  This method, as with the most effective prior research \cite{lagesse_detecting_2018}, works with common computing equipment and still works even if the attacker is using encryption or is on a different Wi-Fi network.  Unlike prior work, this method works both indoors and outdoors without requiring any manipulation of the environment.   

To validate the feasibility of this approach, we first analyzed the significance of the difference of several computationally efficient similarity measurements.  Then, we examined the effectiveness of using those similarity measurements as a threshold-based classifier.  Next, we applied machine learning to further improve our classification results.  As a result, we demonstrated a threshold-based similarity measure that achieved an F1 score of 0.886 and a neural network model that achieved an F1 score of 0.966 with 100\% recall across all of our scenarios.

Next we introduced a signal delay attacker model.  The attacker can delay the streaming of the video.  This attack drastically reduces the effectiveness of the original Similarity of Simultaneous Observation algorithm with very little delay necessary.  To combat this, we introduce additional similarity metrics and use them to train an LSTM model.  The new model not only defeats the delayed streaming attacker, but it also improves the overall performance of the system over our previous work.

From these results, we conclude that Similarity of Simultaneous Observation is an effective approach to detecting hidden streaming cameras in a variety of environments where previous work has failed.  We have identified that there are some environments in which the technique performs better than others, but even in the most difficult environments our work is valuable.

\section{Human Subjects and Ethical Considerations}

The experiments described in this paper were reviewed by our IRB and were determined to be exempt from a full IRB review since any humans that were incidentally captured by our cameras were in public locations and the techniques rely only on the bytes per time step of the recorded video, not the content of the video.

\bibliographystyle{elsarticle-num}
\bibliography{SimilarityOfSimultaneousObservation}

\end{document}